\documentstyle[sprocl]{article}

\def\Journal#1#2#3#4{{#1} {\bf #2}, #3 (#4)}

\def\NPB{{\em Nucl. Phys.} B}
\def\PLB{{\em Phys. Lett.}  B}
\def\PRL{\em Phys. Rev. Lett.}
\def\PRD{{\em Phys. Rev.} D}

\def\be{\begin{equation}}
\def\ee{\end{equation}}
\def\bea{\begin{eqnarray}}
\def\eea{\end{eqnarray}}

\begin{document}

\title{UNITARITY CONSTRAINTS ON NEUTRINO MASS AND MIXINGS}

\author{A. B. BALANTEKIN}

\address{Max-Planck-Institut f\"ur Kernphysik,\\
Postfach 103980, D-69029 Heidelberg, Germany\\
and\\
Department of
Physics, University of Wisconsin,  \\
Madison, Wisconsin 53706
USA\footnote{Permanent address. 
Electronic address: {\tt baha@nucth.physics.wisc.edu}}}


\maketitle\abstracts{We explore the implications of imposing the
  constraint that two 
  neutrino flavors (which for definiteness we take to be $\nu_{\mu}$
  and $\nu_{\tau}$) are similarly coupled to the mass basis in
  addition to the unitarity constraints. Implications of this scheme
  for specific experimental situations are discussed.}

\section{Introduction}

\indent

Recent observations of atmospheric neutrinos and especially their
zenith-angle dependence \cite{atm}, strongly suggest that muon
neutrinos maximally mix with the tau neutrinos. Motivated by this
observation and recent theoretical work on neutrino mass models
\cite{ramond}, we explored \cite{uniter} the implications of imposing
the constraint that two neutrino flavors (which for definiteness we
take to be $\nu_{\mu}$ and $\nu_{\tau}$) are similarly coupled to the
mass basis in addition to the unitarity constraints.

Although the invisible width of the Z particle constraints the number
of active neutrino flavors to be three, it is nevertheless worthwhile
to consider the possibility of the existence of sterile neutrino
states for a number of reasons: i) The possibility of oscillation of
atmospheric muon neutrinos into sterile states is not completely ruled
out. ii) If the LSND results \cite{lsnd} are confirmed, since the
analysis of LSND, atmospheric \cite{atm} and solar \cite{solar}
neutrinos point out to different mass scales, one needs to introduce
sterile neutrinos. iii) Serious problems such as the abundance of
alpha particles that arise when core-collapse supernovae with
neutrino-driven wind are considered as sites of r-process
nucleosynthesis can be avoided by the oscillations of active neutrinos
into sterile ones \cite{gail,cfq}. Even though cosmological and
astrophysical bounds rule out heavier sterile states \cite{raffelt},
the effect of the lighter sterile neutrinos on big-bang
nucleosynthesis is controversial \cite{bbn}. 

Hence we consider three active flavors and an  arbitrary number (which
could be taken to be zero) of sterile neutrinos. The $N \times N$
neutrino mixing matrix will be denoted by $U_{\alpha i}$ where
$\alpha$ denotes the flavor index and $i$ denotes the mass index:
\begin{equation}
  \label{eq:1} | \nu_{\alpha} \rangle = \sum_i U_{\alpha i} | \nu_i
  \rangle.
\end{equation}
We impose the constraint that $U_{\mu i}$ and $U_{\tau i}$ are
proportional for all but one mass eigenstate, which we choose for
definiteness to be the third mass eigenstate:
\begin{equation}
  \label{eq:2} U_{\mu i} \sim  U_{\tau i} \neq 0, \> \forall i \neq 3.
\end{equation}
We write this condition in terms of an arbitrary angle $\phi$ and an
arbitrary phase $\eta$:
\begin{equation}
  \label{eq:3} \sin \phi \> U_{\mu i} =  e^{i\eta} \cos \phi \>
  U_{\tau i} \neq 0, \> \forall i \neq 3.
\end{equation}
Note that, in our formalism, we permit CP-violating phases.
Introducing the quantity
\begin{equation}
  \label{4} A = \sum_{i \neq 3} \left[ |U_{\mu i}|^2 + |U_{\tau i}|^2
  \right]
\end{equation}
and using Eq. (\ref{eq:3}) along with the unitarity of the mixing
matrix one can easily show that \cite{uniter}
\begin{equation}
  \label{5} A =1,
\end{equation}
\begin{equation}
  \label{6} U_{\mu 3} = - \sin \phi e^{i\delta} e^{i\eta},
\end{equation}
\begin{equation}
  \label{7} U_{\tau 3} = \cos \phi  e^{i\delta},
\end{equation}
where $\delta$ is a phase to be determined, and 
\begin{equation}
  \label{eq:8} U_{\alpha 3} = 0 , \>\> \alpha \neq \mu,\tau,
\end{equation}
Introducing the states
\begin{equation}
  \label{eq:8a} |  \tilde{\nu}_{\mu} \rangle =  \cos \phi | \nu_{\mu}
  \rangle +  \sin \phi e^{i\eta} | \nu_{\tau} \rangle ,
\end{equation}
and
\begin{equation}
  \label{eq:9} |  \tilde{\nu}_{\tau} \rangle =  - \sin \phi e^{-i\eta}
  | \nu_{\mu} \rangle + \cos \phi | \nu_{\tau} \rangle  ,
\end{equation}
It follows that 
\begin{equation}
  \label{eq:10} | \tilde{\nu}_{\mu} \rangle = \frac{1}{\cos \phi}
  \sum_{i \neq 3} U_{\mu i} | \nu_i \rangle,
\end{equation}
\begin{equation}
  \label{eq:11} | \tilde{\nu}_{\tau} \rangle =  e^{i \delta} | \nu_3
  \rangle.
\end{equation}
and
\begin{equation}
  \label{eq:12} | \nu_{\alpha} \rangle = \sum_{i \neq 3} U_{\alpha i}
  | \nu_i \rangle, \>\> \alpha \neq \mu,\tau.
\end{equation}

This is a remarkable result which simply follows from the assumption
of Eq. (\ref{eq:3}). This assumption leads to a decoupling of
all the other flavors from the chosen (the third in
our choice) mass eigenstate in the neutrino mixing matrix. 
 
\subsection*{Three Active Flavors}

\indent

For three active flavors we get 
\begin{equation}
  \label{eq:13} \left(\matrix{ | \nu_e \rangle \cr | \tilde{\nu}_{\mu}
  \rangle  } \right)~ =  \left(\matrix{ U_{e 1} & U_{e 2}  \cr U_{\mu
  1}/ \cos \phi &  U_{\mu 2}/ \cos \phi  } \right)~  \left(\matrix{ |
  \nu_1 \rangle \cr | \nu_2 \rangle  } \right)~. 
\end{equation}
The solar neutrino data in this case could be explained by either the
matter-enhanced or vacuum $\nu_e \rightarrow \tilde{\nu}_{\mu}$
oscillations.

In the special case of $\phi = \pi/4$, the full mixing matrix is given
by \cite{alter}
\begin{equation}
  \label{eq:14} \left(\matrix{ \cos \theta & - \sin \theta & 0  \cr
  \sqrt{2} \sin \theta & \sqrt{2} \cos \theta &
  \frac{1}{\sqrt{2}}e^{i\delta} \cr \sqrt{2} \sin \theta & \sqrt{2}
  \cos \theta & - \frac{1}{\sqrt{2}}e^{i\delta} \cr } \right)~~.
\end{equation}
The limiting case of  $\theta = \pi/4$ and $\delta = 0$ yields
bi-maximal mixing of three active neutrinos
\cite{bimax1,bimax2,georgig}.

\subsection*{An Arbitrary Number of Flavors}

\indent

In general $N$ flavors mix with the fundamental representation of
$U(N)$. An arbitrary $U(N)$ element can be written as a product of
$N(N-1)/2$ different non-commuting $SU(2)$ rotations and a diagonal
matrix: 
\begin{equation}
  \label{eq:15} U^\dagger_{i\alpha} = R_{12} R_{13} R_{14} \cdots
R_{23} R_{24} \cdots \left( \matrix{ e^{i\delta_1} & 0 & 0 & . \cr 0 &
e^{i\delta_2} & 0 & . \cr 0 & 0 & e^{i\delta_3} & . \cr . & .& .&
. \cr} \right)
\end{equation}
where e.g.
\begin{equation}
  \label{eq:16} R_{14} = \left( \matrix{ C_{14} & 0 & 0 & S^*_{14} &.
\cr 0 & 1 & 0 &0 & . \cr 0 & 0 & 1 &0 & . \cr -S_{14} & 0
&0&C_{14}^*&. \cr . & .& .& . &. \cr} \right) .
\end{equation}
Our choice of parameters leads to
\begin{eqnarray}
\label{concon}
C_{\alpha 3} &=& 1, \forall \alpha \neq 2 \nonumber \\ C_{23} &=& \cos
\phi \nonumber \\ S_{23} &=& e^{i \eta} \sin \phi , 
\end{eqnarray}
hence our choice reduces the number of parameters from $N(N-1)/2$ to
$(N^2 - 3 N + 4 )/2$. 

\section{Specific Cases}

\indent

Here we summarize implications of our scheme for three different
experimental situations. 

\subsection{Atmospheric Neutrinos}

\indent

If we have only active neutrinos with $m_1 \sim m_2$ we have the
standard result: 
\begin{equation}
P(\nu_{\mu} \rightarrow \nu_{\tau}) = \sin^2 2\phi \sin^2 \left[
\frac{(m_3^2 -m_2^2) L}{4E} \right]
\end{equation}
If we have only one sterile state in addition to the active neutrinos
and  there is the mass hierarchy $m_4 > m_3 \gg m_2 > m_1$ (where
$m_2^2 -m_1^2$ is of order of the solar neutrino solution) we get the
following result for the $\nu_{\mu} \rightarrow \nu_{\tau}$ conversion
probability:
\begin{eqnarray}
\label{manyf}
&P&(\nu_{\mu}\rightarrow \nu_{\tau}) = \sin^2 2\phi \sin^2
 \left[\frac{(m_3^2 -m_2^2)L}{4E} \right]  + 4 |U_{\mu 4}|^4 \tan^2
 \phi   \sin^2 \left[\frac{(m_4^2 -m_2^2)L}{4E} \right] \nonumber \\
 &-& 8 \sin^2 \phi |U_{\mu 4}|^2 \sin \left[\frac{(m_4^2 -m_2^2)L}{4E}
 \right]  \sin \left[\frac{(m_3^2 -m_2^2)L}{4E} \right] \cos
 \left[\frac{(m_4^2 -m_3^2)L}{2E} \right] .
\end{eqnarray}
It will be instructive to do a fit to the SuperKamiokande  atmospheric
neutrino data with Eq. (\ref{manyf}). 

\subsection{Reactor Neutrinos}

\indent

In our scheme, if the value of $(m_2^2 -m_1^2)$ is determined from the
solar neutrino data, for reactor neutrino experiments we can  assume
$(m_2^2 -m_1^2) \ll E/L$. We then have
\begin{equation}
P(\nu_e \rightarrow \nu_e) = 1 - \sin^2 2 \theta_{\rm eff} \sin^2
\left[ \frac{(m_4^2 -m_1^2) L}{4E} \right],
\end{equation}
where
\begin{equation}
\sin^2 2 \theta_{\rm eff} = 4 |U_{e 4}|^2 (1 - |U_{e 4}|^2). 
\end{equation}
For the large values of $(m_4^2 -m_1^2)$ that would help the r-process
nucleosynthesis in the neutrino-driven wind models of supernova CHOOZ
experiment gives a bound of $|U_{e 4}|^2 < 0.047$ \cite{chooz}. The
best limit,  $|U_{e 4}|^2 < 0.005$ comes from the BUGEY experiment
\cite{bugey} and is still consistent with the conversion into sterile
neutrinos in supernovae \cite{gail}. 

\subsection{Neutrinoless Double Beta Decay}

\indent

The current data indicates 
\begin{equation}
{\cal M}_{ee} = \sum_i m_i |U_{ei}|^2 < \sim 0.5 {\rm eV}.
\end{equation}
In our scheme $ {\cal M}_{ee} = \sum_{i \neq 3} m_i |U_{ei}|^2$. Thus
for three flavors $ {\cal M}_{ee}$ depends only on $m_1$ and $m_2$,
not on $m_3$.It is possible to enforce  $ {\cal M}_{ee} \equiv 0$ for
bi-maximal mixing \cite{bimax1,georgig}. When sterile neutrinos are
included this puts a limit on $m_4$. One should emphasize that the
uncertainties of the nuclear matrix elements could be rather large so
it may not be necessary to impose  $ {\cal M}_{ee} \equiv 0$. 

\section{Conclusions}

\noindent 

We explored the implications of imposing the constraint that two
neutrino flavors (which for definiteness we take to be $\nu_{\mu}$ and
$\nu_{\tau}$) are similarly coupled to the mass basis in addition to
the unitarity constraints. We allow three active and an arbitrary
number of sterile neutrinos. We show that in this scheme one of the
mass eigenstates decouples from the problem, reducing the dimension of
the flavor space by one. This result allows significant simplification
in the treatment of matter-enhanced neutrino transformation where
multiple flavors and level crossings are involved. 

When the constraint of Eq. (\ref{concon}) is imposed, which was
motivated by the recent experimental results at Superkamiokande, the
form of Eq. (\ref{eq:15}) indicates the existence of a coset structure
of the neutrino mixing matrix. Recent related work discussed the
existence of a an $Sp(4)$ symmetry in the neutrino mass sector
\cite{nurcan}. It was shown that   the most general neutrino mass
Hamiltonian sits in the $Sp(4)/SU(2)\times U(1)$ coset space where
U(1) is the chirality transformation and the $SU(2)$ generates the
see-saw transformation.  At the moment it is not clear what the 
relation, if any, between these two coset structures is.

\section*{Acknowledgments}

\noindent

I thank G. Fuller and T. Weiler for discussions. 
This work was supported in part by the U.S. National Science
Foundation Grant No.\ PHY-9605140 at the University of Wisconsin, 
in part by the University of Wisconsin Research Committee with funds
granted by the Wisconsin Alumni Research Foundation, and in part by 
the Alexander von Humboldt-Stiftung, Germany. The very kind
hospitality of Hans Weidenm\"uller  at the Max-Planck-Institut f\"ur
Kernphysik is much appreciated.

\end{document}